\documentclass{svjour3}                     
\smartqed  
\usepackage{graphicx}
 \usepackage{mathptmx}      
\newcommand{\qq}{Q'} 
\newcommand{\re}{\mathrm{Re}\,}

\newcommand{\gev}{\,{\rm GeV}}

\def\pb{\,{\rm pb}}

\newcommand{\AmS}{{\protect\the\textfont2
  A\kern-.1667em\lower.5ex\hbox{M}\kern-.125emS}}

\begin{document}

\title{ On timelike Compton scattering at medium and high energies  \thanks{Presented at the workshop "30 years of strong interactions", Spa, Belgium, 6-8 April 2011.}}


\author{B. Pire         \and L. Szymanowski \and J. Wagner       
}


\institute{B. Pire \at
             CPHT,  Ecole Polytechnique,  CNRS, Palaiseau, France   \\
              \email{pire@cpht.polytechnique.fr}           
           \and
           L. Szymanowski and J. Wagner \at Soltan Institute for Nuclear Studies, Warsaw, Poland \\
              \email{lechszym@fuw.edu.pl,  kubawag@yahoo.com}              
}

\date{Received: date / Accepted: date}

\maketitle

\begin{abstract}
We emphasize the complementarity of timelike and spacelike studies of deep exclusive processes, 
taking as an example the case of timelike Compton Scattering (TCS) i.e. 
the exclusive photoproduction of a lepton pair 
with large invariant mass, vs deeply virtual Compton scattering (DVCS) i.e. 
the exclusive leptoproduction of a real photon. Both amplitudes factorize with the same 
generalized parton distributions (GPDs) as their soft parts and coefficient functions which 
differ significantly at next to leading order in $\alpha_s$. We also stress that  data on 
TCS at very high energy should be available soon thanks to the study of ultraperipheral 
collisions at the LHC, opening a window on quark and gluon GPDs at very small skewness.
\keywords{generalized parton distributions \and exclusive reactions \and photoproduction}
 \PACS{13.60.Fz \and13.90.+i}
\end{abstract}

\section{Introduction}
\label{intro}
The study of the deep structure of the nucleon has been the subject of many developments 
in the past decades and the concept of generalized parton distributions has allowed a 
breakthrough in the 3 dimensional description  of the quark and gluon content of hadrons. 
Deep exclusive reactions have been demonstrated to allow to probe the 
quark and gluon content of protons and heavier nuclei. 
Results on DVCS \cite{DVCSexp} obtained at HERA and JLab already allow to get a rough 
idea of some of the GPDs (more precisely on Compton form factors) in a restricted but 
interesting kinematical domain \cite{fitting}. An extended research program at JLab@12 GeV and 
Compass is now proposed to go beyond this first set of analysis. 
This will involve taking into account  next to leading order in $\alpha_s$ and next to 
leading twist contributions \cite{APT}.

\section{TCS vs dVCS}
\label{sec:1}
A considerable amount of theoretical and experimental work has 
been devoted to the study of deeply virtual Compton scattering (DVCS),
 i.e., $\gamma^* p \to \gamma p$, 
an exclusive reaction where generalized parton
distributions (GPDs) factorize from perturbatively calculable coefficient functions, when
the virtuality of the incoming photon is high enough~\cite{historyofDVCS}.
It is now recognized that the measurement of GPDs should contribute in a decisive way to
our understanding of how quarks and gluons assemble into
hadrons~\cite{gpdrev}. In particular the transverse
location of quarks and gluons become experimentally measurable via the transverse momentum dependence of the GPDs \cite{Burk}.

  The physical process where to observe the inverse reaction,  timelike Compton scattering (TCS) \cite{BDP},
 \begin{equation}
 \gamma(q) N(p) \to \gamma^*(q') N(p')
 \end{equation}
 is  (see Fig. 1) the exclusive photoproduction of a
heavy lepton pair, $\gamma N \to \mu^+\!\mu^-\, N$ or $\gamma N \to
e^+\!e^-\, N$, at small $t = (p'-p)^2$ and large \emph{timelike} virtuality $q'^2 = Q'^2$ of the final state photon; TCS 
shares many features with DVCS. The Bjorken variable in that case is $\tau = Q'^2/s $
 with $s=(p+q)^2$. One also defines $\Delta = p' -p$  ($t= \Delta^2$) and the skewness variable 
  $\eta$ as
$$\eta = - \frac{(q-q')\cdot (q+q')}{(p+p')\cdot (q+q')} \,\approx\,
           \frac{ Q'^2}{2s  - Q'^2} = \frac{ \tau}{ (2-\tau)}.$$
  $x$ and $\eta$ represent plus-momentum fractions 
  $$x = \frac{(k+k')^+}{(p+p')^+} , \;
\eta \approx  \frac{(p-p')^+}{(p+p')^+} .$$
At the Born order, the TCS amplitude is described by the handbag diagrams of Fig. 2.
\begin{figure}
 \hspace*{4cm} \includegraphics[width=6cm]{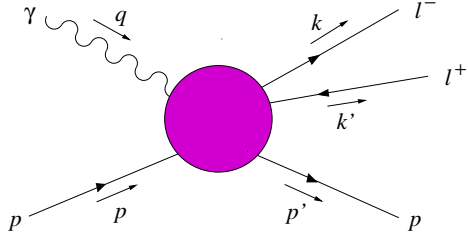}
\caption{Real photon-proton scattering into a lepton pair
and a proton.}
\label{fig:1}       
\end{figure}

 As in the case of DVCS, a purely electromagnetic
mechanism where the lepton pair is produced through the Bethe-Heitler (BH) subprocess (see Fig. 3)
$$\gamma (q)\gamma^* (\Delta) \to\ell^+\ell^-\;,$$  contributes at the amplitude level. This amplitude 
is completely calculable in QED provided one knows the  nucleon form factors at small $\Delta^2 = t$.
This process has a very peculiar angular dependence and overdominates the TCS process if
one blindly integrates over the final phase space. One may however choose kinematics where 
the amplitudes of the two processes are of the same order of magnitude, and either subtract the 
well-known Bethe-Heitler process or use specific observables sensitive to
 the interference of the two amplitudes. Finally some kinematical cuts may allow to decrease 
 sufficiently the Bethe Heitler contribution.
 
\begin{figure*}
  \includegraphics[width=0.5\textwidth]{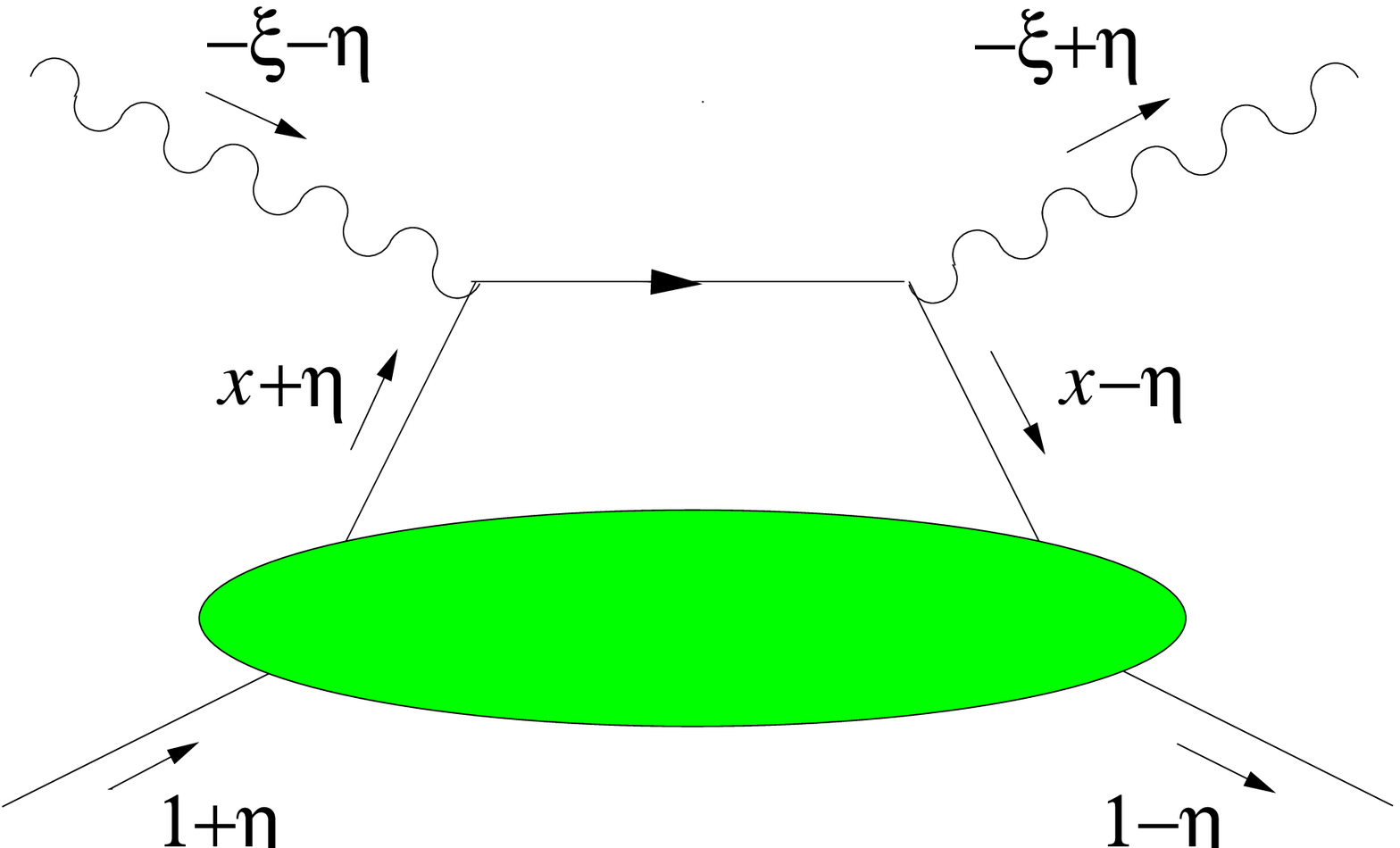}
\hspace*{0cm} \includegraphics[width=0.5\textwidth]{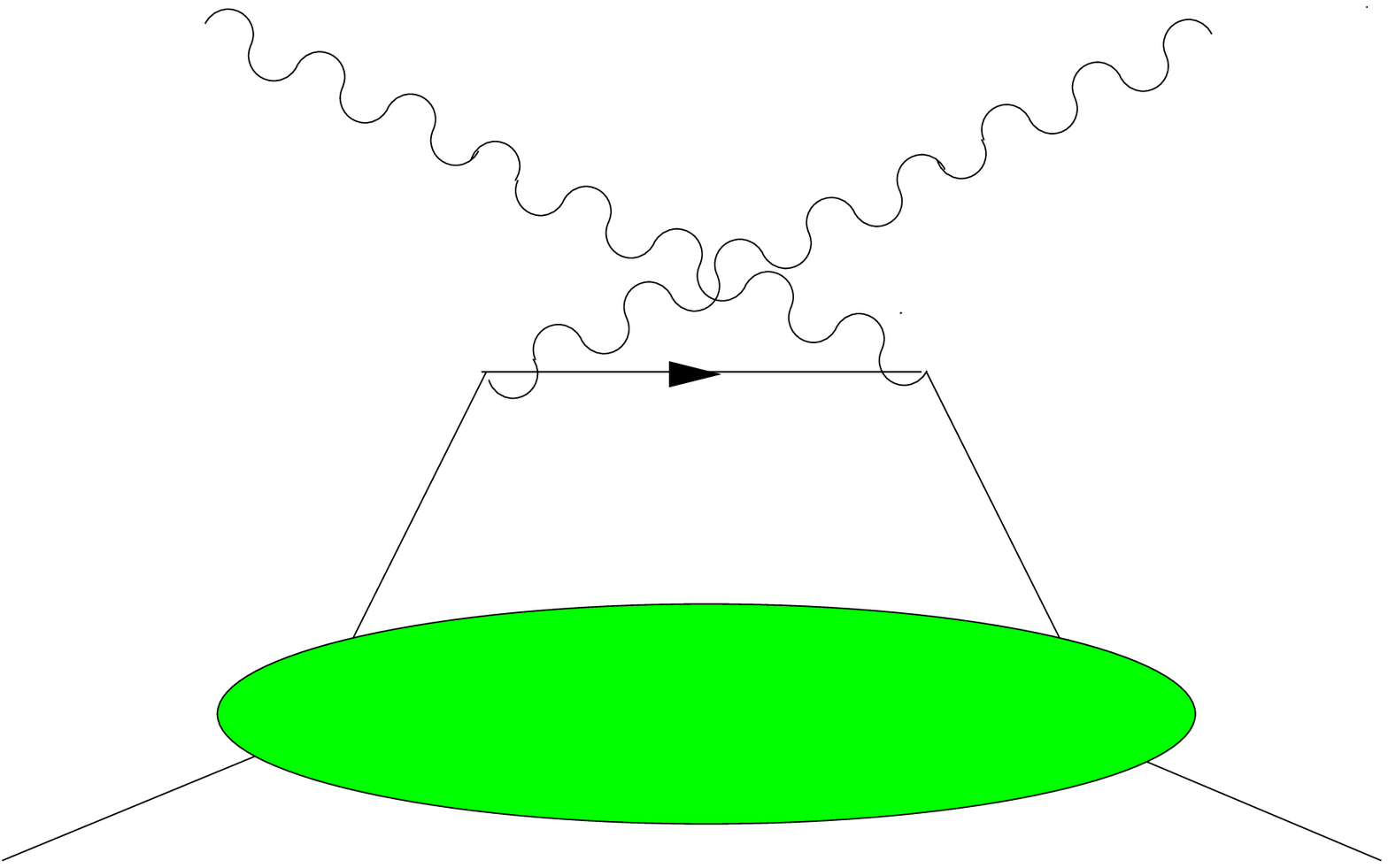} 
\caption{Handbag diagrams for the Compton process in the scaling limit. The
plus-momentum fractions $x$, $\xi$, $\eta$ refer to the average proton
momentum $\frac{1}{2}(p+p')$. In the DVCS case, $\xi = \eta$ while in the TCS case $\xi = -\eta$.}
\label{fig:2}       
\end{figure*}

\begin{figure*}
\hspace*{0cm}  \includegraphics[width=1\textwidth]{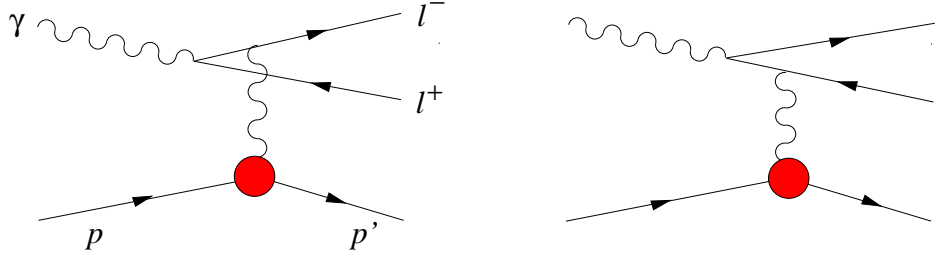}
\caption{The Feynman diagrams for the Bethe-Heitler amplitude.}
\label{fig:3}       
\end{figure*}

\begin{figure}
 \hspace*{0cm} \includegraphics[width=12cm]{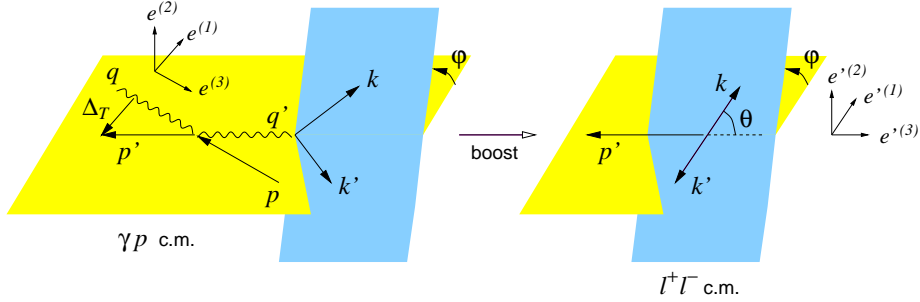}
\caption{Kinematical variables and coordinate axes in
the $\gamma p$ and $\ell^+\ell^-$ c.m.\ frames.}
\label{angle}      
\end{figure}

The kinematics of the $\gamma (q) N (p)\to \ell^-(k) \ell^+(k') N(p')$ process is shown in Fig. \ref{angle}.
 In the $\ell^+\ell^-$ center of mass system,  one introduces the polar and azimuthal angles $\theta$
and $\varphi$ of $\vec{k}$, with reference to a coordinate system with
$3$-axis along $-\vec{p}\,'$ and $1$- and $2$-axes such that $\vec{p}$
lies in the $1$-$3$ plane and has a positive
$1$-component.

 This program has not yet been experimentally successful \cite{NadelTuronski:2009zz} due to the existing limited quasi real photon flux in the right kinematical domain both at JLab and HERA. This will be  much improved with the JLab@12 GeV program, both in Hall B \cite{Albayrak} and in Hall D. These experiments will enable to test the universality of GPDs extracted from DVCS and from TCS, provided NLO corrections are taken into account. Experiments at higher energies, e.g. in ultraperipheral collisions at RHIC and LHC \cite{PSW}, may even become sensitive to gluon GPDs which enter the amplitude only at NLO level.

\section{TCS in  ultraperipheral reactions}
\label{sec:2}

We estimated the different contributions to the lepton pair cross section for 
ultraperipheral collisions at the LHC.  Since the
cross sections decrease rapidly with $Q'^2$, we are interested in the kinematics of moderate $Q'^2$, 
say a few GeV$^2$, and large energy, thus very small values of $\eta$. 
Note however that for a given proton energy the photon flux is higher at 
smaller photon energy.

\vskip.1in
$\bullet$ The Bethe Heitler cross section

\begin{figure*}{h}
  \includegraphics[width=0.5\textwidth]{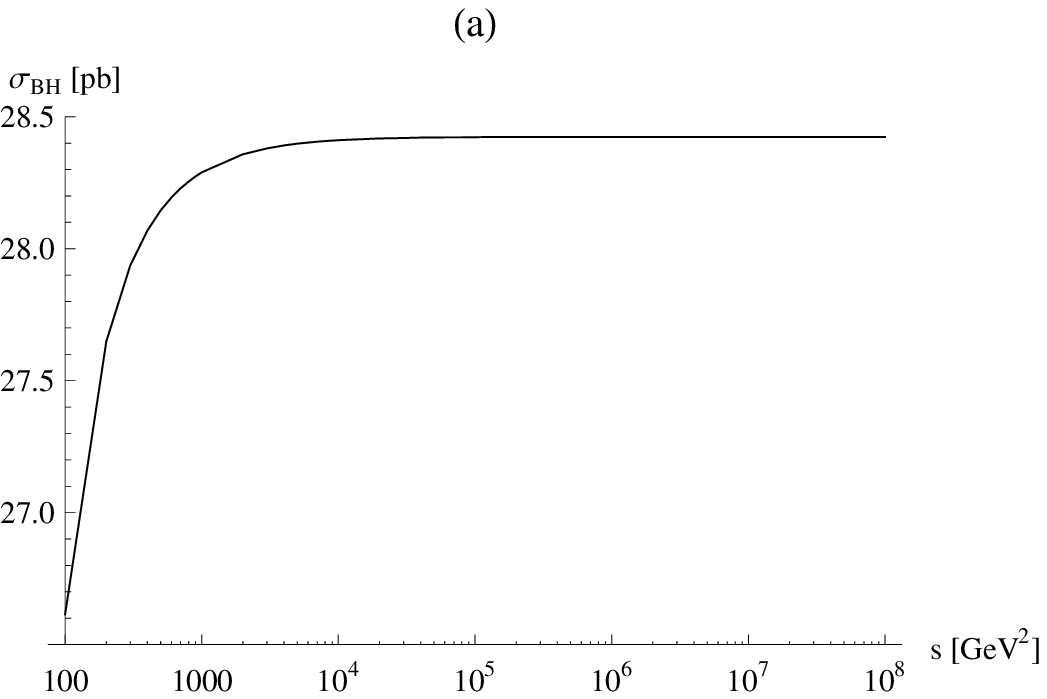}
\includegraphics[width=0.5\textwidth]{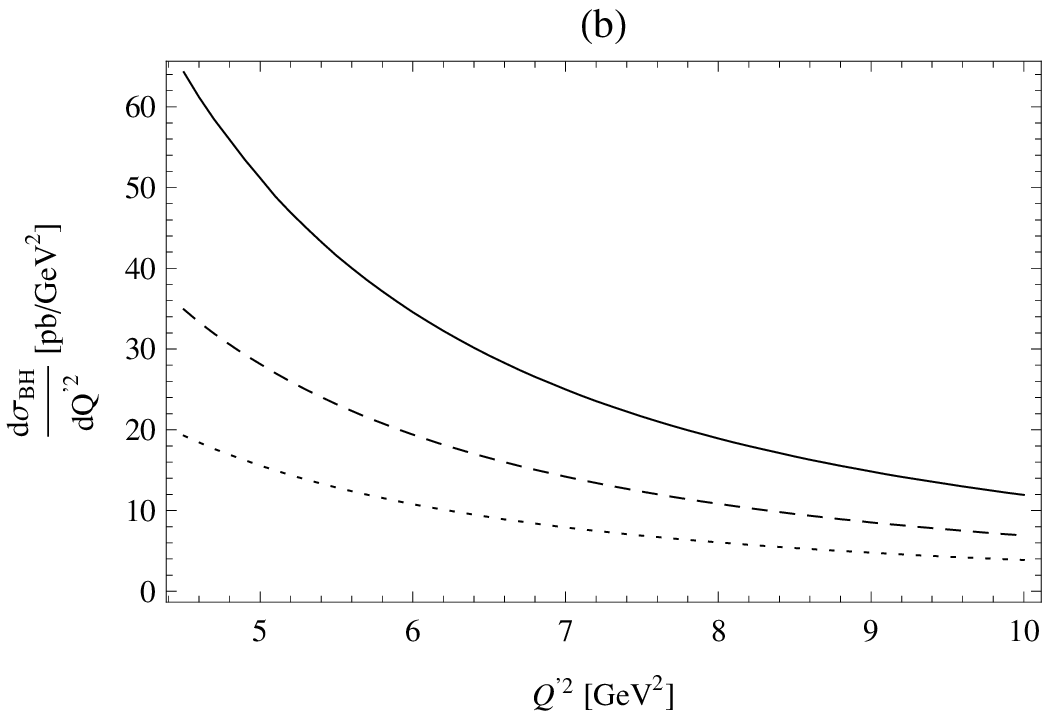} 
\caption{(a) The BH cross section integrated over $\theta \in [\pi/4,3
\pi/4]$, $\varphi \in [0, 2\pi]$ , $Q'^2 \in [4.5,5.5]\gev^2$, $|t| \in [0.05,0.25] \gev^2$, as a function of $\gamma p$ c.m. energy squared $s$.
(b) The BH cross section integrated over $\varphi \in [0, 2\pi]$ , $|t| \in [0.05,0.25] \gev^2$, and various ranges of 
$\theta$ : $[\pi/3,2 \pi/3]$ (dotted), $[\pi/4,3 \pi/4]$ (dashed) 
and $[\pi/6,5 \pi/6]$ (solid), as a function of ${Q'}^2$ for $s=10^5 \gev^2$}
\label{BHS}       
\end{figure*}

\noindent
 The full Bethe Heitler cross section integrated over $\theta \in [\pi/4,3
\pi/4]$, $\varphi \in [0, 2\pi]$ , $Q'^2 \in [4.5,5.5]\gev^2$, $|t| \in [0.05,0.25] \gev^2$, as a function of $\gamma p$  energy squared $s$ is shown on Fig. \ref{BHS}. We see that in the limit of large
 $s$ it is constant and equals $28.4 \pb$. 
On Fig. \ref{BHS}b, the Bethe Heitler contribution is shown as a function of $Q'^2$ when it is integrated over  $\varphi$ in the range $[0, 2 \pi]$, $-t $ in the range $[0.05, 0.25]$GeV$^2$ and for  $\theta$ integrated  in various ranges $[\pi/3,2 \pi/3]$, $[\pi/4,3 \pi/4]$ and $[\pi/6,5 \pi/6]$. 
As anticipated, the cross section grows much when small $\theta$ angles are allowed.
In the following we will use the limits $[\pi/4,3 \pi/4]$ where the cross section is sufficiently big but does not dominate too much over the Compton process.


\vskip.1in
$\bullet$ The TCS cross section

\begin{figure*}{h}
  \includegraphics[width=0.5\textwidth]{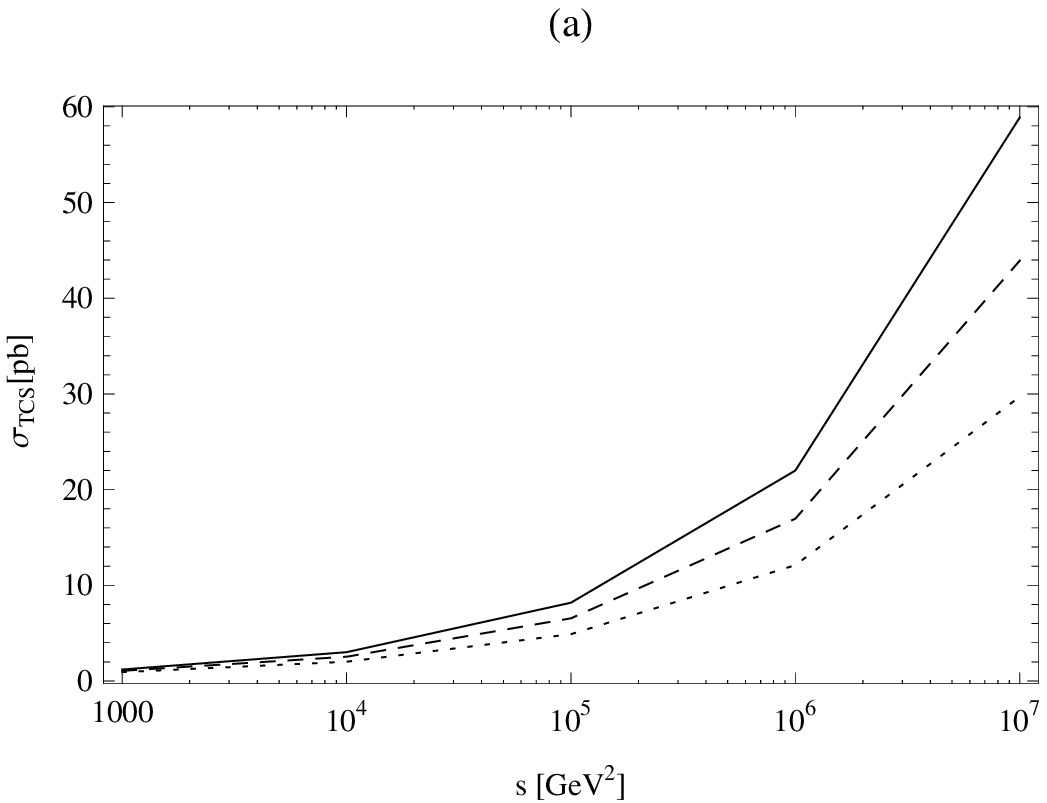}
\includegraphics[width=0.5\textwidth]{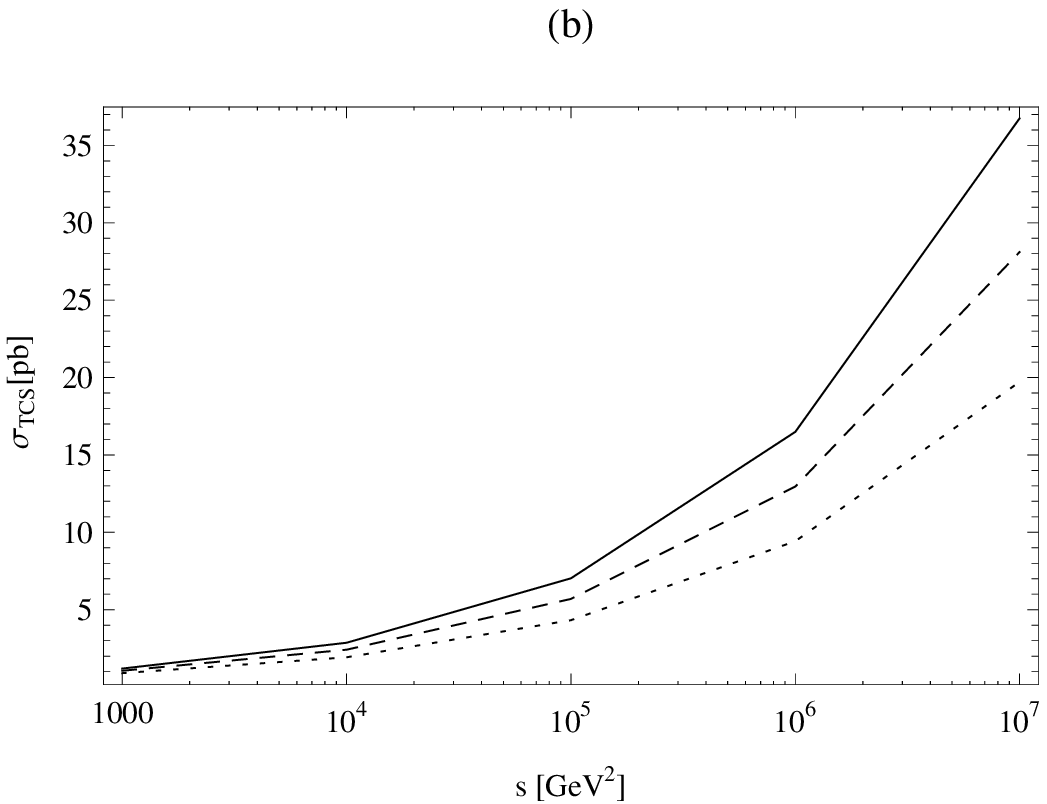} 
\caption{$\sigma_{TCS}$ as a function of $\gamma p$ c.m. energy squared $s$, for 
GRVGJR2008 LO (a) and NLO (b) parametrizations, for different factorization scales 
$\mu_F^2 = 4$ (dotted), $5$ (dashed), $6$ (solid) $\gev^2$.}
\label{Sigma_TCS}
\end{figure*}

\noindent
On Fig. \ref{Sigma_TCS} we plot the leading order Compton cross section $\sigma_{TCS}$  as a function of the  photon-proton energy  squared $s$. For very high energies $\sigma_{TCS}$ calculated with $\mu_F^2 = 6 \gev^2$ is much bigger then with $\mu_F^2 = 4\gev^2$. Also predictions obtained using LO and NLO GRVGJR2008 PDFs differ significantly. 
 
\vskip.1in
$\bullet$ The interference cross section


\noindent
Since the amplitudes for the Compton and Bethe-Heitler
processes transform with opposite signs under reversal of the lepton
charge,  the interference term between TCS and BH is
odd under exchange of the $\ell^+$ and $\ell^-$ momenta.
It is thus possible to
 project out
the interference term through a clever use of
 the angular distribution of the lepton pair. 
The interference part of the cross-section for $\gamma p\to \ell^+\ell^-\, p$ with 
unpolarized protons and photons has a characteristic ($\theta, \varphi$) dependence given  by (see details in \cite{PSW})
\begin{eqnarray}
   \label{intres}
\frac{d \sigma_{INT}}{d\qq^2\, dt\, d\cos\theta\, d\varphi}
= {}-
\frac{\alpha^3_{em}}{4\pi s^2}\, \frac{1}{-t}\, \frac{M}{Q'}\,
\frac{1}{\tau \sqrt{1-\tau}}\,
  \cos\varphi \frac{1+\cos^2\theta}{\sin\theta}
     \re{M} \; ,\nonumber
\end{eqnarray}
with 
\begin{equation}
\label{mmimi}
{M} = \frac{2\sqrt{t_0-t}}{M}\, \frac{1-\eta}{1+\eta}\,
\left[ F_1 {\cal H}_1 - \eta (F_1+F_2)\, \tilde{\cal H}_1 -
\frac{t}{4M^2} \, F_2\, {\cal E}_1 \,\right],
\nonumber
\end{equation}
where $-t_0 = 4\eta^2 M^2 /(1-\eta^2)$ and ${\cal H},  \tilde{\cal H}, {\cal E}$ are Compton form factors.
 With the integration limits symmetric about $\theta=\pi/2$ the interference
term changes sign under $\varphi\to \pi+\varphi$ due to charge conjugation,
whereas the TCS and BH cross sections do not. One may thus extract the 
Compton amplitude through a study of
$\int\limits_0^{2\pi}d\phi\,\cos \phi \frac{d\sigma}{d\phi}$.

\begin{figure*}
  \includegraphics[width=0.5\textwidth]{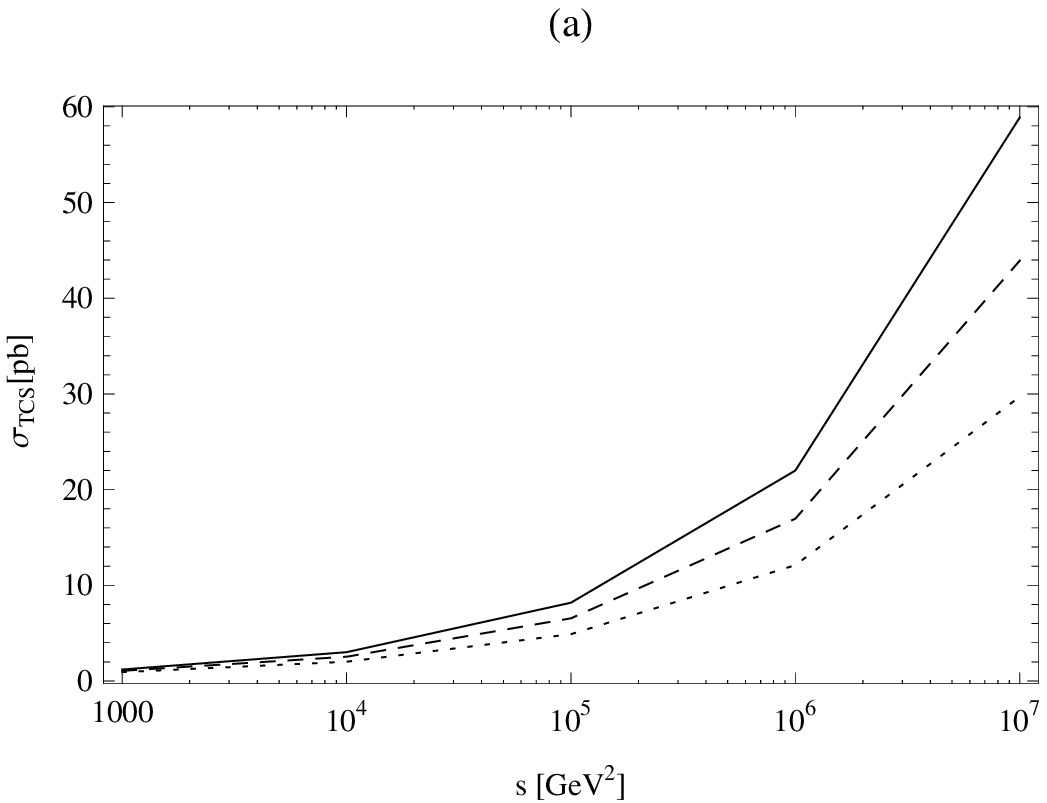}
\includegraphics[width=0.5\textwidth]{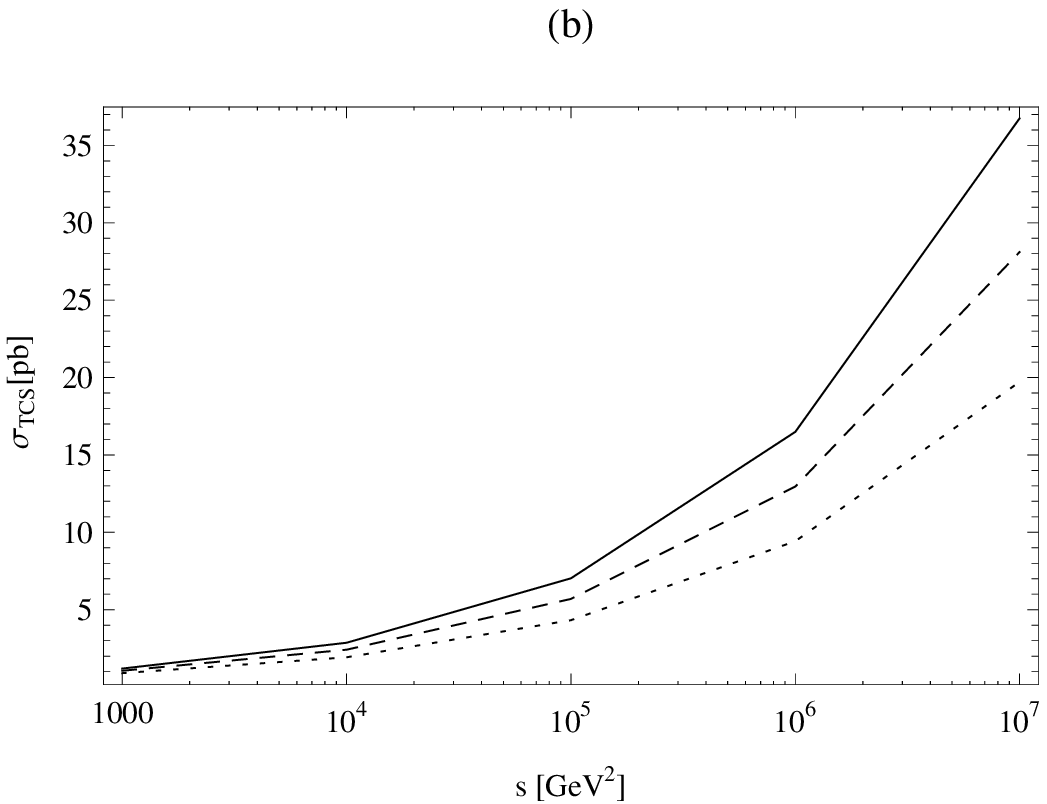} 
\caption{$\sigma_{TCS}$ as a function of $\gamma p$ c.m. energy squared $s$, for 
GRVGJR2008 LO (a) and NLO (b) parametrizations, for different factorization scales 
$\mu_F^2 = 4$ (dotted), $5$ (dashed), $6$ (solid) $\gev^2$.}
\label{Sigma_TCS}
\end{figure*}

\begin{figure*}
  \includegraphics[width=0.5\textwidth]{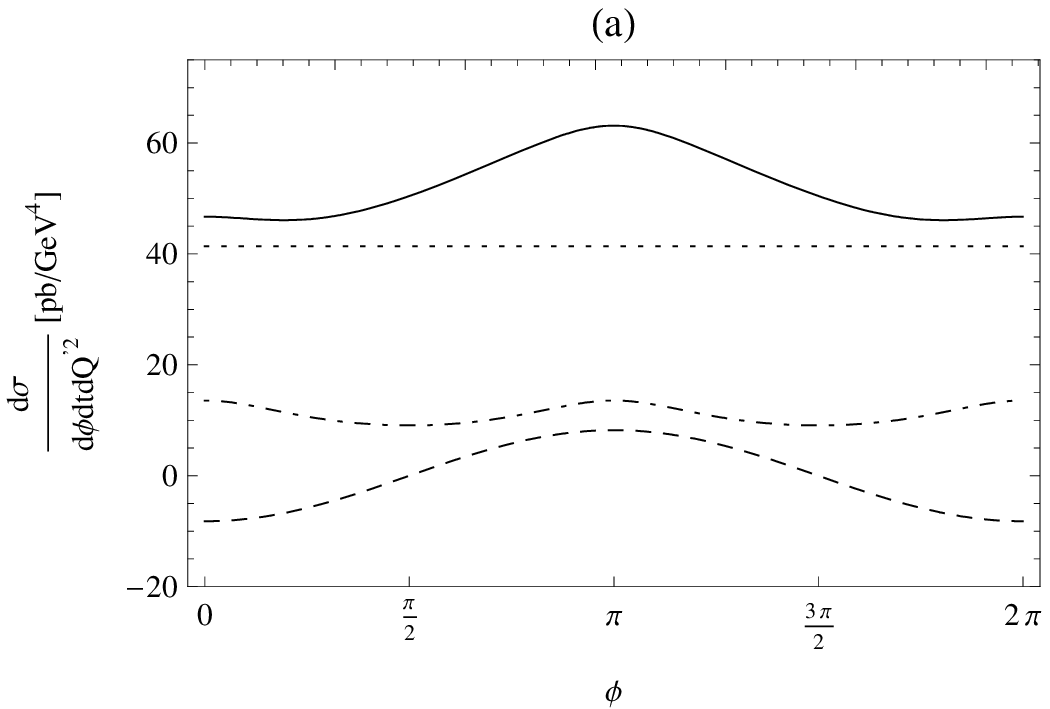}
\includegraphics[width=0.5\textwidth]{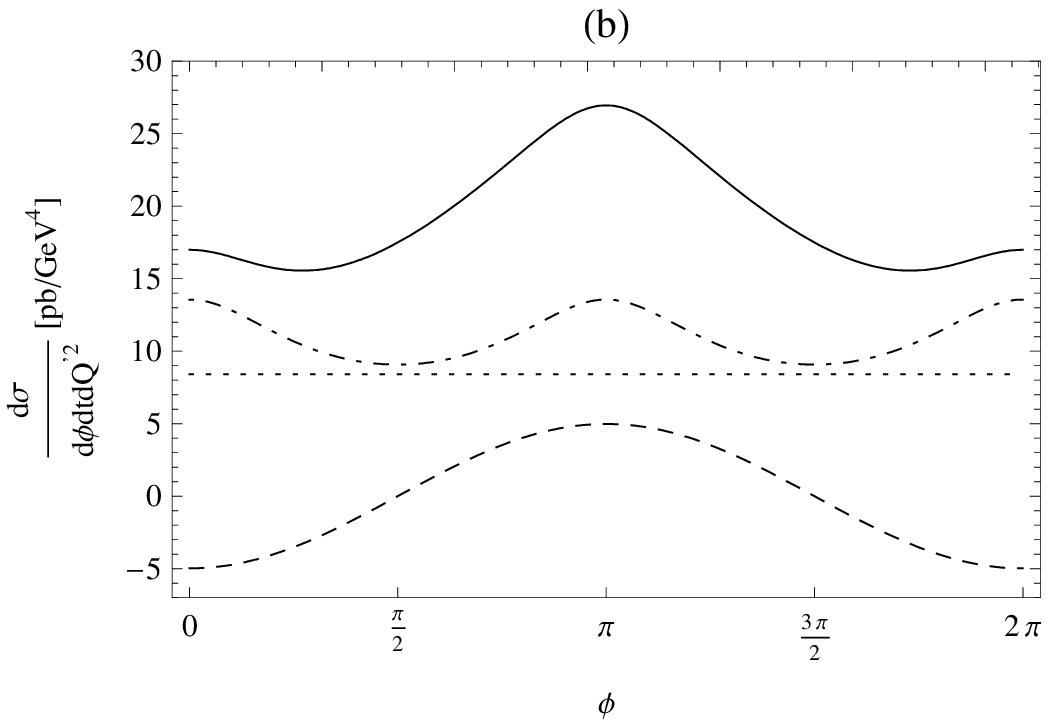} 
\caption{The differential cross sections (solid lines) for $t =-0.2 \gev^2$, ${Q'}^2 =5 \gev^2$  integrated 
over $\theta = [\pi/4,3\pi/4]$, as a function of $\varphi$, for $s=10^7 \gev^2$ (a), 
$s=10^5 \gev^2$(b)
 with $\mu_F^2 = 5 \gev^2$. We also display  the
Compton (dotted), Bethe-Heitler (dash-dotted) and Interference (dashed) contributions.}
\label{Interf}
\end{figure*}

In Fig. \ref{Interf} we show the interference contribution to the cross section in comparison to the Bethe Heitler and Compton processes, for various values of photon proton  energy 
squared $s = 10^7 \gev^2,10^5 \gev^2$. We observe that for 
larger energies the  
Compton process dominates, whereas for $s=10^5 \gev^2$ all contributions are comparable.

\vskip.1in
\noindent
In conclusion, timelike Compton scattering in ultraperipheral collisions at hadron 
colliders opens a new way to measure generalized parton distributions, in particular for very small values of the skewness parameter. Our leading  order estimate show that the factorization scale dependence of the amplitudes is quite high. This fact demands the understanding of higher order contributions with the hope that they will stabilize this scale dependence.

\section{NLO corrections}
\label{sec:3}
TCS and DVCS amplitudes are identical (up to a complex conjugation) at lowest order in $\alpha_S$ but differ at next to leading order, in particular because of the quite different analytic structure of these reactions. Indeed the production of a timelike photon enables the production of intermediate states in some channels which were kinematically forbidden in the DVCS case. This opens the way to new absorptive parts of the amplitude.  

Our calculations \cite{PSW2} are performed along the lines of \cite{JiO} (see also \cite{NLO2}). We shall not repeat here the results but focus on 
 important differences between the coefficient functions describing the TCS case and  those describing DVCS. First, the $p^2 +i\varepsilon$ prescription for propagators turns into a  $\eta \to \eta+i\varepsilon$, rather then a  $\eta \to \eta-i\varepsilon$ as in the DVCS case. The second difference is the presence of minus signs under the logarithms, which  produce additional terms. Particularly $\log^2(-2-i\varepsilon)$ present in the TCS result may produce correction much bigger then the corresponding $\log^2(2)$ in the DVCS case. Another important difference between the DVCS and TCS amplitudes appear in their imaginary part, which  is present only in the DGLAP region for DVCS, while it is present in both DGLAP and ERBL regions for TCS. Defining the quark and gluon coefficient functions as
 \begin{eqnarray}
T^q = C_0^q+ C_1^q +\frac{1}{2}ln(\frac{ |Q^2|}{\mu_F^2}) C_{coll}^q ~~~~~; ~~~~~
T^g =  C_1^g +\frac{1}{2}ln(\frac{ |Q^2|}{\mu_F^2}) C_{coll}^g\;,
\nonumber
\label{eq:NLOTCSDVCS}
\end{eqnarray}
where $ C_0^q$ is the Born order coefficient function and $\mu_F$ is the factorization scale. $C^q_{coll} $ and $C^g_{coll} $ are directly related to the evolution equation kernels.

\begin{figure*}
  \includegraphics[width=0.5\textwidth]{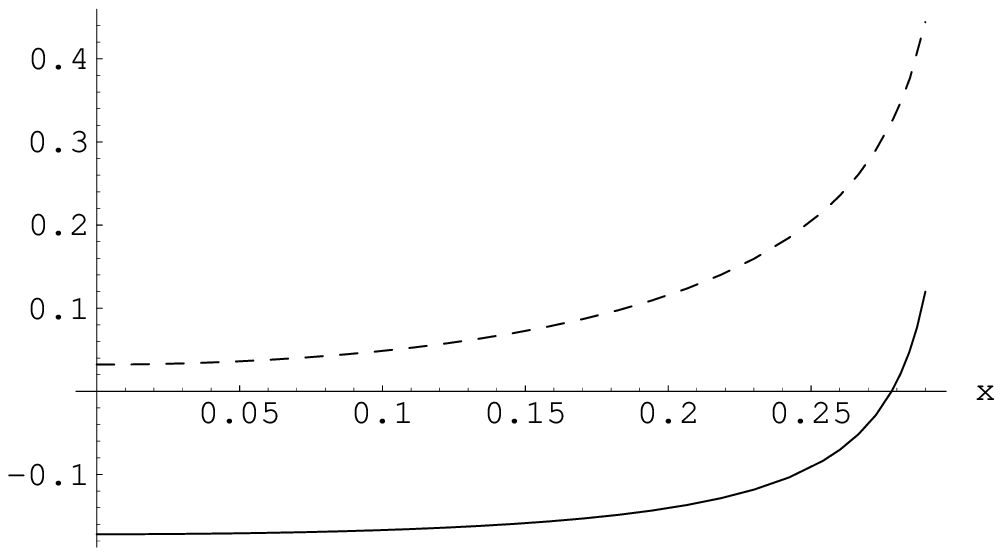}
\includegraphics[width=0.5\textwidth]{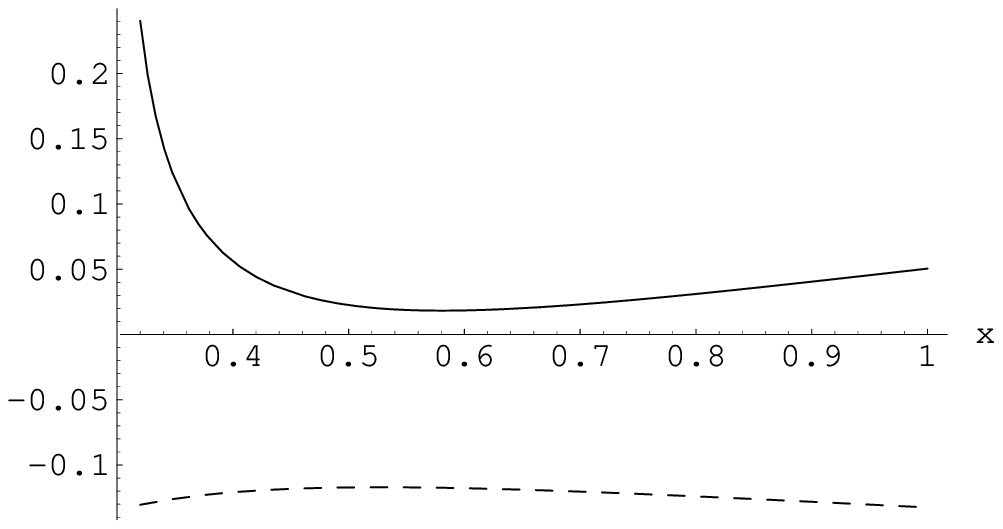} 
\includegraphics[width=0.5\textwidth]{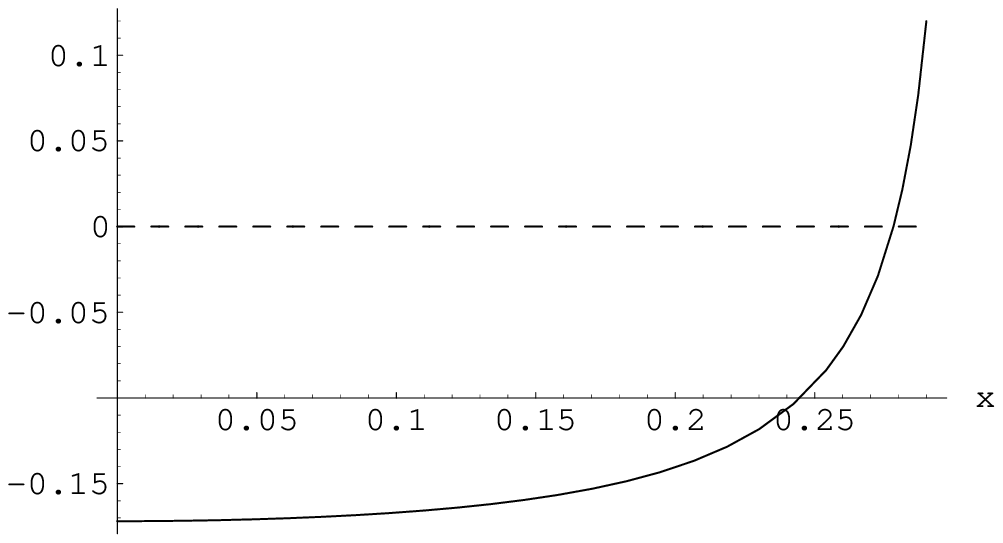} 
\includegraphics[width=0.5\textwidth]{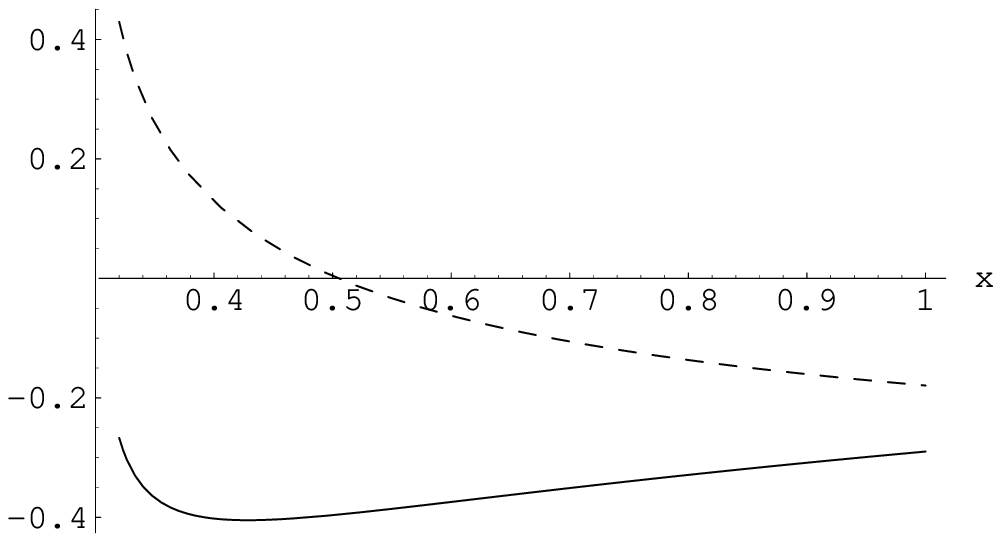} 
\caption{Real (solid line) and imaginary (dashed line) part of the ratio $R^q$ of the NLO quark coefficient function to the Born term in Timelike Compton Scattering (up) and Deeply Virtual Compton Scattering (down) as a function of $x$ in the ERBL (left) and DGLAP (right) region for $\eta = 0.3$, for $\mu_F^2 = |Q^2|$.}
\label{Fig:ratio}
\end{figure*}

To discuss the difference of the coefficient functions ${C_{1(TCS)}^q}^* - C_{1(DVCS)}^q  $ and present the magnitude of corrections we define the following ratio:
\begin{eqnarray}
R^q = \frac{C_{1}^q+\frac{1}{2}\log \left(\frac{|Q^2|}{\mu_F^2}\right) \cdot C^q_{coll}}{C^q_{0}}
\label{eq:r}
\end{eqnarray}
of the NLO quark correction to the coefficient function, to the Born level one. Let us restrict us to the factorization scale choice $\mu_F^2 = |Q^2|$.
On Fig. \ref{Fig:ratio} we  show  the real and imaginary parts of the ratio $R^q$ in timelike and spacelike Compton Scattering as a function of $x$ in the ERBL (left) and DGLAP (right) region for $\eta = 0.3$. We  fix $\alpha_s = 0.25$ and restrict the plots to the positive $x$ region, as the coefficient functions are antisymmetric in that variable. We see that in the TCS case, the imaginary part of the amplitude is  present in both the ERBL and DGLAP regions, contrarily to the DVCS case, where it exists  only in the DGLAP region. The magnitudes of these NLO coefficient functions are not small. We see that the importance of these NLO coefficient functions is magnified when we consider the  difference of the coefficient functions  ${C_{1(TCS)}^q}^* - C_{1(DVCS)}^q  $. The conclusion is that extracting the universal GPDs from both TCS and DVCS reactions requires much care.

\begin{figure*}
\hspace*{3cm} \includegraphics[width=0.5\textwidth]{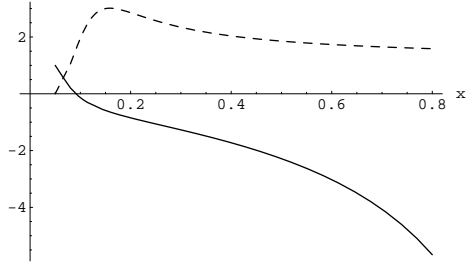}
\caption{Ratio of the real (solid line) and imaginary (dashed line) part of the NLO gluon coefficient function in TCS to the same quantity in DVCS as a function of $x$ in the DGLAP  region for $\eta = 0.05$ for $\mu_F^2 = |Q^2|$.
}
\label{Fig:ratiog}
\end{figure*}

Let us now briefly comment on  the gluon coefficient functions. 
The  real parts of the gluon contribution are equal for DVCS and TCS in the ERBL region. 
The differences between TCS and DVCS emerges in the ERBL region through the imaginary part of the coefficient function which is non zero only for the TCS case and is of the order of the real part. 
In Fig. \ref{Fig:ratiog} we plot the ratio 
$\frac{C^g_{1(TCS)} }{ C^g_{1(DVCS)}}$
of the NLO gluon correction to the hard scattering amplitude in TCS to the same quantity in the DVCS in the DGLAP  region for $\eta = 0.05$.

More phenomenological studies need now to be performed, by convoluting the coefficient functions to realistic quark and gluon GPDs and calculating the relevant observables in various kinematical domains. We are now  progressing on these points.

\begin{acknowledgements}
We aknowledge useful discussions with Grzegorz Grzelak, Pawe{\l} Nadel-Turo\'nski and Samuel Wallon. This work was supported by the French-Polish Scientific Agreement Polonium.
\end{acknowledgements}



\end{document}